# Systematic approach to correlators in $T\bar{T}$ deformed CFTs

Song He[◌],[1,2,3,4,*] Yuan Sun[◌],[5,†] and Jiashi Yin[◌][6,7,‡]

[1]*Institute of Fundamental Physics and Quantum Technology,*
*Ningbo University, Ningbo, Zhejiang 315211, China*
[2]*School of Physical Science and Technology, Ningbo University, Ningbo, Zhejiang 315211, China*
[3]*Center for Theoretical Physics and College of Physics, Jilin University, Changchun 130012, China*
[4]*Max Planck Institute for Gravitational Physics (Albert Einstein Institute),*
*Am Mühlenberg 1, 14476 Golm, Germany*
[5]*Institute of Quantum Physics, School of Physics, Central South University, Changsha 418003, China*
[6]*Beijing No. 4 High School, Beijing 100032, China*
[7]*Department of Physics, University of Illinois, Urbana, Illinois 61801-3080, USA*

We investigate higher-order corrections to correlators in a general CFT (conformal field theory) with the double-trace $T\bar{T}$ deformation. Standard perturbation theory proves inadequate for this problem due to the intricate stress-tensor flow induced by the deformation. To tackle this challenge, we introduce a novel technique termed the conservation equation method. This method leverages the stress tensor's trace relation and conservation property to establish relationships between higher and lower-order corrections, and subsequently determine the correlators by enforcing symmetry properties. As an illustration, we compute first- and higher-order corrections to various types of correlators. Our results align with existing calculations in the literature and demonstrate the viability of our method for general CFTs.

## I. INTRODUCTION

The $T\bar{T}$ deformation of two-dimensional quantum field theories, as introduced by Smirnov *et al.* [1–3], has garnered significant attention owing to its remarkable properties. This deformation offers improved analytic tractability compared to generic irrelevant deformations. Several studies have demonstrated its integrability [3–7]. Under the $T\bar{T}$ flow, the deformed spectrum remains exactly solvable, preserving an infinite tower of conserved charges and their associated algebra. Various equivalent descriptions have been provided, including those in terms of string theory [8,9], a random metric problem [10,11], and 2D gravity [12,13]. Moreover, accumulating evidence suggests a potential link between deformed CFTs (conformal field theories) and cutoff-AdS (Anti-de Sitter) gravity, offering a novel example of holography beyond conventional holographic CFTs [14–21]. An alternative holographic description involves imposing a mixed boundary condition at the asymptotic AdS boundary [22,23].

Considerable efforts have been directed toward computing correlators in $T\bar{T}$-deformed CFTs through diverse methodologies. Noteworthy contributions on the partition function, or the zero-point function, include works [24–26]. These studies explore the deformed spectrum and delve into the modular properties of the deformed partition function. Prior investigations have computed one-point functions of Korteweg-de Vries charges on a torus [27,28]. Perturbative calculations to the first nontrivial order in the deformation parameter $\lambda$ encompass higher-point functions of the stress tensor [11,15,20,29–31] and undeformed operators [32–37]. Additionally, surface charges in $T\bar{T}$-deformed theories have been calculated to impose constraints on correlators [38–40]. The renormalized Lagrangians of the deformed massive boson and Dirac fermion have been constructed by leveraging integrability [4,41].

Notably, there have been several nonperturbative investigations, covering aspects such as UV divergences of correlators [42] and the large-momentum behavior of two-point correlators [43]. A recent functional renormalization group study on $T\bar{T}$-deformed scalar field theory has unveiled a nontrivial UV fixed point [44]. Other relevant studies include a nonperturbative computation of two-point

[*]Contact author: hesong@nbu.edu.cn
[†]Contact author: sunyuan@csu.edu.cn
[‡]Contact author: jacob.yin@icloud.com

correlators within the context of the TsT/$T\bar{T}$ correspondence [45,46].

The $T\bar{T}$-deformation of a general 2D quantum field theory (QFT) can be defined by the following flow equation for the action[1]:

$$\partial_\lambda S^\lambda = \frac{1}{\pi} \int d^2x \mathcal{O}_{T\bar{T}}(x), \qquad (1)$$

where $\mathcal{O}_{T\bar{T}} \equiv T\bar{T} - \Theta^2$, $T \equiv -2\pi T_{zz}$, and $\Theta \equiv 2\pi T_{z\bar{z}} = \frac{\pi}{2}T^\mu_\mu$. Here, $T_{\mu\nu}$ denotes the stress tensor defined in the deformed theory, which introduces a nonlinear dependence on $\lambda$. A corresponding flow equation for correlators [2] reads[2]

$$\begin{aligned}\partial_\lambda \left\langle \prod_i O_i(z_i) \right\rangle^\lambda &= \left\langle \partial_\lambda \left( \prod_i O_i(z_i) \right) \right\rangle^\lambda \\ &\quad - \frac{1}{\pi} \int d^2x \left\langle \mathcal{O}_{T\bar{T}}(z) \prod_i O_i(z_i) \right\rangle^\lambda \\ &\quad - \langle \mathcal{O}_{T\bar{T}}(z)\rangle^\lambda \left\langle \prod_i O_i(z_i) \right\rangle^\lambda. \end{aligned} \qquad (2)$$

The first term on the rhs of (2) captures the flow effect of deformed operators whose functional forms depend on $\lambda$, such as conserved currents (e.g., the stress tensor operator). From a Hamiltonian perspective, these deformed operators differ from their CFT counterparts even on the initial time slice, evolving not only with the deformed Hamiltonian but also undergoing changes in their explicit forms [47]. Undeformed operators do not receive explicit $\lambda$-dependent corrections to their functional forms, so the first term above vanishes in that case. Since the disconnected term vanishes on a Euclidean plane (see Appendix A), it is omitted.

Expanding both sides of the equation in powers of $\lambda$ yields a relationship between higher-order and lower-order corrections,

$$\begin{aligned}\left\langle \prod_i O_i(z_i)\right\rangle^{(n)} &= \sum_{m=0}^{n-1} \left\langle \left(\prod_i O_i(z_i)\right)^{(m)}\right\rangle^{(n-m)} \\ &\quad - \frac{1}{n\pi} \int d^2x \left\langle \mathcal{O}_{T\bar{T}}(z)\prod_i O_i(z_i)\right\rangle^{(n-1)},\end{aligned} \qquad (3)$$

where $A^{(n)}$ represents the coefficient of order $\lambda^n$ in the series expansion of a given quantity within the deformed theory, expressed as $A^\lambda = \sum_i \lambda^i A^{(i)}$. Specifically, $A^{(0)}$ corresponds to the operator in the CFT. For example, $\langle \prod_i O_i(z_i)\rangle^{(n)}$ represents the $n$-th-order correction to the correlator $\langle \prod_i O_i(z_i)\rangle^\lambda$, and $O^{(n)}$ represents the $n$-th-order correction to the functional form of the deformed operator $O$ in terms of the fundamental fields. As an illustration, for the deformed massless free boson, the deformed stress tensor is found to be [48]

$$\mathcal{O}_{T\bar{T}} = \frac{4\lambda(\partial\phi\bar{\partial}\phi)^2 + \sqrt{1 - 8\lambda(\partial\phi\bar{\partial}\phi)^2} - 1}{2\lambda^2\sqrt{1 - 8\lambda(\partial\phi\bar{\partial}\phi)^2}}, \qquad (4)$$

$$\mathcal{O}_{T\bar{T}} = \sum_i \lambda^i \mathcal{O}^{(i)}_{T\bar{T}}, \qquad \mathcal{O}^{(0)}_{T\bar{T}} = (2\pi)^2(\partial\phi\bar{\partial}\phi)^2,$$

$$\mathcal{O}^{(1)}_{T\bar{T}} = 32\pi^3(\partial\phi\bar{\partial}\phi)^3. \qquad (5)$$

Equation (3) shows that higher-order corrections depend on knowing the expansion of the deformed stress tensor. However, for a generic 2D CFT neither an explicit deformed Lagrangian nor its corresponding stress tensor is available in closed form. This difficulty has previously restricted perturbative calculations primarily to special seed theories such as free bosons[3] [28].

In this paper, we address the perturbative calculation of $T\bar{T}$-deformed correlation functions for a general CFT, without relying on an explicit form of the deformed stress tensor. Our approach leverages a recursion relation among higher- and lower-order corrections to stress tensor correlators, combined with bootstrap-like constraints. In one application, we derive the first-order correction $\langle \mathcal{O}_{T\bar{T}}(z)X\rangle^{(1)}$ for a general CFT and thereby obtain second-order corrections to a wide range of correlators. With this result at hand, we naturally recover the known logarithmic behavior at higher orders [42] and isolate contributions from the less tractable terms in $\mathcal{O}_{T\bar{T}}$. Our method also facilitates the calculation of pure stress tensor correlators at orders extending beyond previous analyses.

The paper is organized as follows. In Sec. II, we detail the general procedure for order-by-order computation of deformed correlators. Section III applies these methods to derive first-order corrections for various classes of correlators, including mixed correlators of stress tensors and undeformed operators. Section IV proceeds to second-order results, verifying consistency with known results. We conclude with a summary and possible future directions.

## II. PRESCRIPTION FOR PERTURBATIVE CALCULATION

In this section, we outline a systematic method for computing deformed correlators to arbitrary order. We consider the undeformed theory $S^{\lambda=0}$ to be a conformal field theory on a Euclidean plane. From the flow equation of action (1), one can show that the stress tensor follows the trace relation,

$$\Theta = \lambda \mathcal{O}_{T\bar{T}}, \tag{6}$$

whose validity has been demonstrated for the deformed free boson in [3] and proven by variational principle in [23].

---

[1]Note that our definition differs from that in [15] by an overall factor of $1/\pi$, i.e., $\frac{1}{\pi}(\partial_\lambda S^\lambda)_{\text{ours}} = (\partial_\lambda S^\lambda)_{\text{KLM}}$. Also, $T_{zz}$ in [15] corresponds to $T$ in our paper.
[2]We use complex coordinates, with the holomorphic field coordinate explicitly written, omitting the antiholomorphic dependence for simplicity.

[3]For recent progress in this direction, see [49].

We treat the trace relation (6) as an operator identity valid inside correlation functions,

$$\langle \Theta(z)X \rangle^{\lambda} = \lambda \langle \mathcal{O}_{T\bar{T}}(z)X \rangle^{\lambda}. \tag{7}$$

Expanding both sides in powers of $\lambda$ gives

$$\langle \Theta(z)X \rangle^{(n)} = \langle \mathcal{O}_{T\bar{T}}(z)X \rangle^{(n-1)}, \quad n \neq 0. \tag{8}$$

Furthermore, using the stress tensor conservation equations, we obtain

$$\begin{aligned} \langle T(z)X \rangle^{(n)} &= \langle (\partial_{\bar{z}}^{-1} \partial_z \Theta(z))X \rangle^{(n)}, \\ \langle \bar{T}(z)X \rangle^{(n)} &= \langle (\partial_z^{-1} \partial_{\bar{z}} \Theta(z))X \rangle^{(n)}, \end{aligned} \tag{9}$$

which formally allows us to replace all insertions of $T$ and $\bar{T}$ inside correlators by $\partial_{\bar{z}}^{-1} \partial_z \Theta$ and $\partial_z^{-1} \partial_{\bar{z}} \Theta$. Here $\partial_{\bar{z}}^{-1}$ formally denotes the inverse of $\partial_{\bar{z}}$. Although these antiderivatives are defined only up to holomorphic or antiholomorphic terms, these terms can often be fixed by symmetries or other properties of the correlators in a deformed CFT, as we shall demonstrate in subsequent sections.

With the trace relation and conservation equations in hand, the key insight is that higher-order corrections can be expressed in terms of lower-order ones whenever stress tensor insertions are present within the correlator in question. For correlators of undeformed operators, the stress tensor can be introduced via the flow equation (3).

To compute the $n$-th-order correction to a bare undeformed operator correlator $\langle X \rangle^{(n)}$ in practice, we apply the following steps:

(1) Insert the interaction vertex $\int d^2z \mathcal{O}_{T\bar{T}}$,

$$\langle X \rangle^{(n)} = -\frac{1}{n\pi} \int d^2x \langle \mathcal{O}_{T\bar{T}}(z)X \rangle^{(n-1)}. \tag{10}$$

(2) Express the stress tensor components $T$ or $\bar{T}$ in terms of $\Theta$ via the conservation equations,

$$\int d^2x \langle (T\bar{T} - (\Theta)^2)(z)X \rangle^{(n-1)}$$
$$= \int d^2x \langle (\partial_{\bar{z}}^{-1} \partial_z \Theta \cdot \bar{T} - (\Theta)^2)(z)X \rangle^{(n-1)}. \tag{11}$$

(3) Insert the trace relation to again reduce the order. Hence (11) becomes

$$\int d^2x \langle ((\partial_{\bar{z}}^{-1} \partial_z \mathcal{O}_{T\bar{T}} \cdot \bar{T} - \Theta \mathcal{O}_{T\bar{T}})(z)X \rangle^{(n-2)}, \tag{12}$$

where we adopt the convention that the operators within the operator products are normally ordered successively from right to left.

(4) Repeat until the resulting expression involves only undeformed correlators, i.e., zeroth-order terms.

(5) Perform any necessary antiderivatives and determine the (anti)holomorphic terms through symmetry constraints or specific properties of $\langle X \rangle^{\lambda}$.

This procedure, which we refer to as the conservation equation method, avoids relying on the explicit form of the deformed stress tensor.

To make the expression (12) well defined and finite, point-splitting regularization is applied to all coincident operators. All divergences are then removed by constructing suitable $\lambda$-dependent renormalized operators, order by order.

Before examining examples, we remark on relevant details of the operator $\mathcal{O}_{T\bar{T}}$, which is defined as [1]

$$\lim_{z' \to z} T(z)\bar{T}(z') - \Theta(z)\Theta(z') = \mathcal{O}_{T\bar{T}}(z) + \text{derivative terms}. \tag{13}$$

where total derivative terms on the right-hand side reflect an inherent ambiguity. Such ambiguities do not affect the deformed action itself but can introduce uncertainties in deformed correlators [50]. Consequently, the quantum trace relation may also be subject to ambiguities,

$$\Theta = \lambda \mathcal{O}_{T\bar{T}} + \partial_\mu W^\mu. \tag{14}$$

These "improvement terms" remain unresolved in a general setting. In our analysis, we work in a gauge where such total derivative terms vanish, defining the $T\bar{T}$ operator as

$$\mathcal{O}_{T\bar{T}}(z) = \lim_{z' \to z} T(z)\bar{T}(z') - \Theta(z)\Theta(z'). \tag{15}$$

## III. FIRST-ORDER CORRECTIONS

### A. Correlators of undeformed operators

In this section, we apply the formalism established in the previous section to derive the first-order correction for $T\bar{T}$-deformed correlators involving undeformed operators. We assume that the undeformed ($\lambda = 0$) theory is a conformal field theory on a Euclidean plane. A similar approach was adopted in a previous investigation [42]. Our findings reveal that the first-order correction to a correlator of undeformed operators can be expressed as a sum of correlators involving descendant operators.

At first order, the calculation proceeds straightforwardly. The first-order correction to an arbitrary bare correlator $\langle X \rangle^\lambda = \langle O_1(z_1) \cdots O_N(z_N) \rangle^\lambda$ is given by

$$\langle X \rangle^{(1)} = -\frac{1}{\pi} \int d^2x \langle \mathcal{O}_{T\bar{T}}(z) X \rangle^{(0)}. \qquad (16)$$

Introducing a point-splitting regulator $\varepsilon$, we obtain

$$\langle X \rangle^{(1)} = -\frac{1}{\pi}\lim_{\varepsilon \to 0} \int d^2x \langle (T(z+\varepsilon)\bar{T}(z) - \Theta(z+\varepsilon)\Theta(z))X \rangle^{(0)}. \qquad (17)$$

Using the conformal Ward identity, one obtains

$$\langle X \rangle^{(1)} = -\frac{1}{\pi} \int d^2x \sum_{m,n} \sum_{r,s \geq 1} \left\langle \frac{L_{r-2,m}\bar{L}_{s-2,n}}{(z-z_m+\varepsilon)^r(\bar{z}-\bar{z}_n)^s} X \right\rangle^{(0)}. \qquad (18)$$

Here, $L_{r-2,m}X$ denotes the operator obtained from acting with the Virasoro generator $L_{r-2}$ on the $m$-th insertion (i.e., $[L_{r-2}(z_m), \mathcal{O}(z_m)]$) while leaving the other insertions in $X$ unchanged. The indices $m, n$ run over all insertions, and $r, s$ run over all positive integers. Details of the calculations are provided in Appendix C 1, leading to

$$\langle X \rangle^{(1)} = \langle \mathrm{d}X \rangle^{(0)}, \qquad \mathrm{d} \equiv \sum_{m,n} \mathrm{d}_{z_m,z_n}. \qquad (19)$$

The operator $\mathrm{d}_{z_m,z_n}$ has two cases depending on whether $m \neq n$ or $m = n$

$$\mathrm{d}_{z_m,z_n} X \equiv \begin{cases} \log(|z_{mn}|^2/\varepsilon^2)\partial_{z_m}\partial_{\bar{z}_n} - \sum_{s \geq 2} \frac{1}{s-1} \frac{\bar{L}_{s-2,n}\partial_{z_m}}{\bar{z}_{mn}^{s-1}} - \sum_{r \geq 2} \frac{1}{r-1} \frac{L_{r-2,m}\partial_{\bar{z}_n}}{z_{nm}^{r-1}}, & \text{if } m \neq n, \\ \sum_{s \geq 2} \frac{1}{s-1} \frac{\bar{L}_{s-2,m}\partial_{z_m}}{\varepsilon^{s-1}} - \sum_{r \geq 2} \frac{1}{r-1} \frac{L_{r-2,m}\partial_{\bar{z}_m}}{\varepsilon^{r-1}}, & \text{if } m = n. \end{cases} \qquad (20)$$

Note that this operator is not symmetric under the interchange of $z_m$ and $z_n$.

The expression (20) exhibits both logarithmic and power divergences, which can be removed by introducing suitably renormalized fields as

$$\mathcal{O}_R(z_i) \equiv \mathcal{O}(z_i) + \frac{\lambda}{\pi} \int d^2x \left( \frac{[L_{-1}(z_i),[\bar{L}_{-1}(z_i),\mathcal{O}(z_i)]]}{(z-z_i+\mu)(\bar{z}-\bar{z}_i)} + \sum_{r,s \geq 1} \frac{[L_{r-2}(z_i),[\bar{L}_{s-2}(z_i),\mathcal{O}(z_i)]]}{(z-z_i+\varepsilon)^r(\bar{z}-\bar{z}_i)^s} \right) + O(\lambda^2), \qquad (21)$$

or explicitly,

$$\mathcal{O}_R(z_i) \equiv \mathcal{O}(z_i) - \lambda \left( \log(\mu^2\varepsilon^2)[L_{-1}(z_i),[\bar{L}_{-1}(z_i),\mathcal{O}(z_i)]] \right.$$
$$\left. - \sum_{r,s \geq 2} \frac{1}{s-1} \frac{[L_{-1}(z_i),[\bar{L}_{s-2}(z_i),\mathcal{O}(z_i)]] + [L_{r-2}(z_i),[\bar{L}_{-1}(z_i),\mathcal{O}(z_i)]]}{\varepsilon^{s-1}} \right) + O(\lambda^2), \qquad (22)$$

where $\mu$ is an arbitrary renormalization scale with mass dimension one to make the argument of the logarithm dimensionless. This renormalization scheme is first introduced in [42]. The first-order correction to a correlator of renormalized fields is given by (20) as

$$\langle X_R \rangle^{(1)} = \langle \mathrm{d}_R X \rangle^{(0)} \qquad (23)$$

with

$$\mathrm{d}_R \equiv \sum_{m \neq n} \left( \log(\mu^2|z_{mn}|^2)\partial_{z_m}\partial_{\bar{z}_n} - \sum_{s \geq 2} \frac{1}{s-1} \frac{\bar{L}_{s-2,n}\partial_{z_m}}{\bar{z}_{mn}^{s-1}} - \sum_{r \geq 2} \frac{1}{r-1} \frac{L_{r-2,m}\partial_{\bar{z}_n}}{z_{nm}^{r-1}} \right). \qquad (24)$$

This formulation is free of any divergences at the cost of introducing an arbitrary renormalization scale.

As a first example, consider the two-point correlator of primary fields with conformal dimensions $(h, \bar{h})$. One finds

$$\langle \mathcal{O}_R(z)\mathcal{O}_R(w)\rangle^{(1)} = \frac{8h\bar{h}(\log(\mu^2|z-w|^2)+1)}{(z-w)^{2h+1}(\bar{z}-\bar{w})^{2\bar{h}+1}}, \quad (25)$$

and absorbing the nonlogarithmic power term by rescaling $\mu \to \mu/\sqrt{e}$ yields

$$\langle \mathcal{O}_R(z)\mathcal{O}_R(w)\rangle^{(1)} = \frac{8h\bar{h}\log(\mu^2|z-w|^2)}{|z-w|^2}\langle \mathcal{O}(z)\mathcal{O}(w)\rangle^{(0)}, \quad (26)$$

which agrees with the result in [11,33].

### B. Stress tensor correlators

We now examine correlators of the deformed stress tensor, using the trace relation and conservation equations discussed earlier.

#### 1. Spin constraint

At first order, the stress tensor two-point functions remain unchanged (see below discussion or [15]); however, a closer look reveals a constraint on the holomorphic terms, which we call the "spin constraint." Arising from the conservation equations, this constraint is crucial for understanding the behavior of the stress tensor correlators at first order. While similar results have been observed in other studies [15], our approach offers more general insights for a wider range of scenarios.

Applying the trace relation, the first-order correction to the two-point function $\langle \Theta(z_1)T(z_2)\rangle$ is found to be zero,

$$\langle \Theta(z_1)T(z_2)\rangle^{(1)} = \langle (T\bar{T} - \Theta^2)(z_1)T(z_2)\rangle^{(1)} = 0. \quad (27)$$

By the conservation equations, the first-order correction $\langle T(z_1)T(z_2)\rangle^{(1)}$ is constrained to be a term holomorphic in $z_1$ (and, symmetrically, in $z_2$). The problem of determining the specific form of such holomorphic terms thus arises.

To proceed, we note that the theory's rotational invariance remains unbroken under the deformation. Revisiting the notion of spin, for a field $\phi_i(x)$ with spin $s_i$, one may define a spin operator [51]

$$S = J(0) - z\partial_z + \bar{z}\partial_{\bar{z}}, \quad (28)$$

when acting on the field. Here, $J(0)$ is the total angular momentum operator that generates rotations around the point $z = 0$.[4] Acting on a correlator $\langle X \rangle$ of several fields, one obtains a condition linking the sum of individual spins $s_i$ to the differential action $(-z\partial_z + \bar{z}\partial_{\bar{z}})$ at each insertion point,

$$\sum_i (-z_i\partial_{z_i} + \bar{z}_i\partial_{\bar{z}_i})\langle X\rangle = \sum_i s_i\langle X\rangle, \quad (29)$$

where $s_i$ represents the spin of the field at $z = z_i$, and the sum $\sum_i s_i$ is identified as the spin of the correlator.

Since rotation symmetry is preserved in the deformed theory, the spin of all fields is unchanged by the deformation, and correlators must therefore retain the same total spin. Since the overall spin of a correlator must be preserved, any correction that alters the total spin is ruled out. We refer to this requirement on deformed correlators as the spin constraint. For instance, the first-order correction derived in (20) satisfies this rule. Schematically, the deformation operator can be written as

$$\sum_m (\partial_{z_m}^{-1}T(z_m))\partial_{\bar{z}_m} + (\partial_{\bar{z}_m}^{-1}\bar{T}(z_m))\partial_{z_m},$$

and it leaves the total spin unchanged at each point $z = z_m$. Although $\partial_{\bar{z}_m}$ lowers the spin by one, the factor $\partial_{z_m}^{-1}T(z_m)$ compensates for that change.

Turning to first-order corrections to stress tensor two-point functions, recall that in 2D QFT the $zz$ component of the stress tensor has spin 2. Its two-point function must therefore have total spin 4. Hence, the full correlator takes the general form

$$\langle T(z_1)T(z_2)\rangle^{\lambda} = \frac{f(\lambda, |z_{12}|, \mu, \varepsilon)}{z_{12}^4},$$

where $f(\cdot)$ is spin neutral. If $\langle T(z_1)T(z_2)\rangle^{(1)}$ were a nonzero holomorphic term, it would necessarily scale like $1/z_{12}^6$, giving a spin of 6 and violating the spin constraint. Therefore, such a contribution must vanish. Analogous reasoning shows that $\langle \bar{T}(z_1)\bar{T}(z_2)\rangle^{(1)}$ and $\langle \bar{T}(z_1)T(z_2)\rangle^{(1)}$ also vanish to first order.

We extend this analysis to three-point functions, in line with [20] but incorporating the spin constraint explicitly. Previous arguments found that $\langle T(z_1)T(z_2)T(z_3)\rangle^{(1)}$ (holomorphic in each $z_i$) could not be determined, but here we see that it must be zero to avoid a mismatch in total spin,

$$\langle \Theta(z_1)\bar{T}(z_2)T(z_3)\rangle^{(1)} = \frac{c^2/4}{z_{12}^4\bar{z}_{13}^4},$$

$$\langle T(z_1)T(z_2)\bar{T}(z_3)\rangle^{(1)} = \frac{c^2/3}{z_{12}^5\bar{z}_{13}^3} + (1\leftrightarrow 2),$$

$$\langle T(z_1)T(z_2)T(z_3)\rangle^{(1)} = 0.$$

Symmetry arguments [e.g., $T(z_1) \leftrightarrow T(z_2)$ interchange] fix holomorphic/antiholomorphic terms. The nonholomorphic correlators respect the correct spin sum, while the purely

---

[4]While $z = 0$ is often chosen as a reference point, the choice is arbitrary. One can appropriately opt for any reference point $z_0$, expressing $S$ as $J(z_0) - (z-z_0)\partial_z + (\bar{z}-\bar{z}_0)\partial_{\bar{z}}$.

holomorphic one would have an incorrect total spin if nonzero.

### 2. Four-point functions

At first order, low-point stress tensor correlators do not exhibit the characteristic logarithmic terms often associated with $T\bar{T}$-deformed theories. To see how such terms emerge, we compute $\langle T(z_1)\bar{T}(z_2)T(z_3)\bar{T}(z_4)\rangle^{(1)}$ using the conservation equation method. We find

$$
\begin{aligned}
\langle T(z_1)\bar{T}(z_2)T(z_3)\bar{T}(z_4)\rangle^{(1)} &= \langle(\partial_{\bar{z}_1}^{-1}\partial_{z_1}\Theta)(z_1)\bar{T}(z_2)T(z_3)\bar{T}(z_4)\rangle^{(1)} \\
&= \partial_{\bar{z}_1}^{-1}\partial_{z_1}\langle\mathcal{O}_{T\bar{T}}(z_1)\bar{T}(z_2)T(z_3)\bar{T}(z_4)\rangle^{(0)} \\
&= \partial_{\bar{z}_1}^{-1}\partial_{z_1}\frac{c}{z_{13}^4}\left(\frac{\partial_{\bar{z}_4}}{\bar{z}_{14}} + \frac{2}{\bar{z}_{14}^2} + (4 \leftrightarrow 2)\right)\frac{c}{\bar{z}_{24}^4} \\
&= \frac{c^2}{z_{13}^4\bar{z}_{24}^4}\left[\left(-\frac{4\log(\mu\bar{z}_{14})}{z_{13}\bar{z}_{24}} + \frac{2}{z_{13}\bar{z}_{14}}\right) + (4 \leftrightarrow 2)\right] + \text{holomorphic in } z_1
\end{aligned}
\tag{30}
$$

with an arbitrary mass scale $\mu$ inserted in the logarithm to make its argument dimensionless, which we identify with the renormalization scale. Such a renormalization scale also appears in stress tensor correlators computed in cutoff 3D gravity [31].

Additional holomorphic terms arising from the antiderivative are strongly constrained by symmetry requirements, including invariance under $z_1 \leftrightarrow z_3$ and $z_2 \leftrightarrow z_4$, and under complex conjugation combined with coordinate interchange. In fact, no further terms are allowed other than a possible factor proportional to $\frac{1}{z_{13}^5\bar{z}_{24}^5}$, which can be absorbed by rescaling $\mu$.

The final result is therefore

$$\langle T(z_1)\bar{T}(z_2)T(z_3)\bar{T}(z_4)\rangle^{(1)} = \frac{c^2}{z_{13}^4\bar{z}_{24}^4}\left[\left(-\frac{4\log(\mu\bar{z}_{14})}{z_{13}\bar{z}_{24}} + \frac{2}{z_{13}\bar{z}_{14}}\right) + (4 \leftrightarrow 2)\right]. \tag{31}$$

This expression is consistent with a direct perturbative check in the deformed massless free boson (see Appendix B 1) and aligns with results from the random geometry approach [11].[5]

### C. Mixed correlators

In this section, we investigate a distinct category of correlators, referred to as mixed correlators, which involve both stress tensors and undeformed operators. They play a significant role in our analysis because higher-order corrections to correlators of undeformed operators can be represented as integrals involving these mixed correlators. Furthermore, they provide insight into how the stress tensor influences undeformed fields, in effect modifying the usual conformal Ward identity.

We begin by examining the first-order correction to the bare mixed correlator $\langle\mathcal{O}_{T\bar{T}}X\rangle^\lambda$, as given by

$$
\begin{aligned}
\langle\mathcal{O}_{T\bar{T}}(z)X\rangle^{(1)} &= \lim_{z'\to z}\langle(T(z)\bar{T}(z') - \Theta(z)\Theta(z'))X\rangle^{(1)} \\
&= \lim_{z'\to z}\langle(\partial_{\bar{z}}^{-1}\partial_z\mathcal{O}_{T\bar{T}}(z)\bar{T}(z') - \mathcal{O}_{T\bar{T}}(z)\Theta(z'))X\rangle^{(0)}.
\end{aligned}
\tag{32}
$$

Using the point-splitting definition of $\mathcal{O}_{T\bar{T}}(z)$ and using the conformal Ward identity leads to

$$
\begin{aligned}
\langle\mathcal{O}_{T\bar{T}}(z)X\rangle^{(1)} &= \lim_{z'\to z}\langle(T(z)\bar{T}(z') - \Theta(z)\Theta(z'))X\rangle^{(1)} = \lim_{z'\to z}\partial_{\bar{z}}^{-1}\partial_z\langle T(z)\bar{T}(z)\bar{T}(z')X\rangle^{(0)} \\
&\to \lim_{z\to z'}\sum_{m,n,i}\sum_{t,r,s\geq 1}\left\langle\frac{\bar{L}_{t-2,i}}{(\bar{z}'-\bar{z}_i)^t}\partial_{\bar{z}}^{-1}\partial_z\frac{L_{r-2,m}\bar{L}_{s-2,n}}{(z-z_m)^r(\bar{z}-\bar{z}_n)^s}X\right\rangle^{(0)},
\end{aligned}
\tag{33}
$$

where the indices $m, n, i$ run over the positions of the operators in $X$. Performing the antiderivative $\partial_{\bar{z}}^{-1}$ yields

---

[5]The conservation equation was applied to $T(z_1)$ in the first step of (30), but this choice is not necessary. One may apply the conservation equation to either $T(z_1), \bar{T}(z_2), T(z_3)$ or $\bar{T}(z_4)$ and follow the aforementioned procedures to arrive at the same result.

$$\langle \mathcal{O}_{T\bar{T}}(z)X\rangle^{(1)} = \sum_{m,n,i}\sum_{t,r\geq 1} \lim_{z'\to z}\left\langle \frac{\bar{L}_{t-2,i}}{(\bar{z}'-\bar{z}_i)^t}\left(-\frac{rL_{r-2,m}\bar{L}_{-1,n}}{(z-z_m)^{r+1}}\log(\mu(\bar{z}-\bar{z}_n))\right)X\right\rangle^{(0)}$$

$$+ \sum_{s\geq 2}\lim_{z'\to z}\left\langle \frac{\bar{L}_{t-2,i}}{(\bar{z}'-\bar{z}_i)^t}\left(\frac{r}{s-1}\frac{L_{r-2,m}\bar{L}_{s-2,n}}{(z-z_m)^{r+1}(\bar{z}-\bar{z}_n)^{s-1}}\right)X\right\rangle^{(0)} + \text{holomorphic in } z \quad (34)$$

with $\mu$ being the renormalization scale.

To fix the holomorphic terms, we enforce the symmetry of $\langle(T(z)\bar{T}(z') - \Theta(z)\Theta(z'))X\rangle^\lambda$ under the replacement of $T \leftrightarrow \bar{T}$ followed by $z \leftrightarrow z'$. To ensure symmetry under $T \leftrightarrow \bar{T}$ (along with $z \leftrightarrow z'$), we add the complex conjugate of all factors (except $X$) with $z$ and $z'$ interchanged, giving

$$\langle \mathcal{O}_{T\bar{T}}(z)X\rangle^{(1)} = \sum_{m,n,i}\sum_{t,r\geq 1} \lim_{z'\to z}\left\langle \frac{\bar{L}_{t-2,i}}{(\bar{z}'-\bar{z}_i)^t}\left(-\frac{rL_{r-2,m}\bar{L}_{-1,n}}{(z-z_m)^{r+1}}\log(\mu^2|z-z_n|^2) - \frac{tL_{r-2,m}L_{-1,n}}{(z-z_m)^r(\bar{z}'-\bar{z}_i)}\log(\mu^2|z'-z_n|^2)\right)X\right\rangle^{(0)}$$

$$+ \sum_{s\geq 2}\lim_{z'\to z}\left\langle \frac{\bar{L}_{t-2,i}}{(\bar{z}'-\bar{z}_i)^t}\left(\frac{r}{s-1}\frac{L_{r-2,m}\bar{L}_{s-2,n}}{(z-z_m)^{r+1}(\bar{z}-\bar{z}_n)^{s-1}} + \frac{t}{s-1}\frac{L_{r-2,m}L_{s-2,n}}{(z-z_m)^r(z'-z_n)^{s-1}(\bar{z}'-\bar{z}_i)}\right)X\right\rangle^{(0)}. \quad (35)$$

Taking the $z' \to z$ limit gives

$$\langle \mathcal{O}_{T\bar{T}}(z)X\rangle^{(1)} = \sum_n \log(\mu^2|z-z_n|^2)(\bar{L}_{-1,n}\partial_z + L_{-1,n}\partial_{\bar{z}})\langle \mathcal{O}_{T\bar{T}}(z)X\rangle^{(0)}$$

$$+ \sum_{r,s\geq 2}\left(\frac{1}{s-1}\frac{\bar{L}_{s-2,n}\partial_z}{(\bar{z}-\bar{z}_n)^{s-1}} + \frac{1}{r-1}\frac{L_{r-2,n}\partial_{\bar{z}}}{(z-z_n)^{r-1}}\right)\langle \mathcal{O}_{T\bar{T}}(z)X\rangle^{(0)} + \text{holomorphic in } z. \quad (36)$$

An additional constraint on the holomorphic terms is the presence of the factor $\frac{\bar{L}_{t-2,i}}{(\bar{z}'-\bar{z}_i)^t}$. The only terms that remain possible take the form

$$\sum_{i,j}\sum_{k,t}\left\langle \frac{L_{k-2,j}\bar{L}_{t-2,i}}{(z-z_j)^k(\bar{z}'-\bar{z}_i)^t}f(X,\{x_n\})\right\rangle^{(0)}. \quad (37)$$

Here $f(X,\{x_n\})$ is spin neutral and has a mass dimension of $n+2$ where $n$ is the mass dimension of $X$. Terms proportional to

$$\sum_{i,j}\sum_{k,t}\left\langle \frac{L_{k-2,j}\bar{L}_{t-2,i}}{(z-z_j)^{k+1}(\bar{z}'-\bar{z}_i)^{t+1}}X\right\rangle^{(0)} \quad (38)$$

are forbidden by the spin constraint. This is because $T(z)$ and $\bar{T}(z')$ are supposed to have spins of $+2$ and $-2$, respectively. Yet the above term can be written as $\sim \langle \partial T(z)\bar{\partial}\bar{T}(z')X\rangle^{(0)}$, in which the fields at $(z,\bar{z})$ and $(z',\bar{z}')$ have a spin of $+3$ and $-3$, respectively, thus violating the spin constraint.

To determine $f(X,\{x_n\})$, we use the Ward identity associated with unbroken translation and rotation symmetry of the deformed theory [42]

$$\langle T_{ij}(\vec{x})X\rangle = \sum_m\left(\frac{(x_j - x_{j,m})\partial_i}{|\vec{x}-\vec{x}_m|^2} + \xi\frac{\epsilon_{jk}(x^k - x_m^k)\epsilon_{ia}\partial^a}{|\vec{x}-\vec{x}_m|^2}\right)\langle X\rangle$$

$$+\cdots, \quad (39)$$

with $\xi = -1$ imposed by symmetry under $i \leftrightarrow j$. In complex coordinates, this becomes

$$\langle T(z)X\rangle = \sum_m \frac{\partial_{z_m}}{z-z_m}\langle X\rangle + \cdots \quad (40)$$

Hence, the only $O(1/(z\bar{z}'))$ term in $\langle T(z)\bar{T}(z')X\rangle^\lambda$ is

$$\sum_{m,n}\frac{\partial_{z_m}\partial_{\bar{z}_n}}{(z-z_m)(\bar{z}'-\bar{z}_n)}\langle X\rangle^\lambda, \quad (41)$$

which leads to a constraint on the $n$-th-order correction,

$$\langle T(z)\bar{T}(z')X\rangle^{(n)} = \sum_{m,n}\frac{\partial_{z_m}\partial_{\bar{z}_n}}{(z-z_m)(\bar{z}-\bar{z}_n)}\langle X\rangle^{(n)} + \cdots \quad (42)$$

where the ellipsis denotes non-$O(1/(z\bar{z}'))$ terms that may or may not be singular. Consequently, we identify $f(X,\{x_n\})$ with $dX$ from (19), ensuring the presence of a term

$$\sum_{i,j} \frac{\partial_{z_i} \partial_{\bar{z}_j}}{(z-z_j)(\bar{z}'-\bar{z}_i)} \langle dX \rangle^{(0)} = \sum_{i,j} \frac{\partial_{z_i} \partial_{\bar{z}_j}}{(z-z_j)(\bar{z}'-\bar{z}_i)} \langle X \rangle^{(1)}. \tag{43}$$

Thus, the full expression reads

$$\langle \mathcal{O}_{T\bar{T}}(z)X \rangle^{(1)} = \sum_n \log(\mu^2 |z-z_n|^2)(\partial_z \partial_{\bar{z}_n} + \partial_{\bar{z}} \partial_{z_n}) \langle \mathcal{O}_{T\bar{T}}(z)X \rangle^{(0)}$$
$$+ \sum_{r,s \geq 2} \left( \frac{1}{s-1} \frac{\bar{L}_{s-2,n} \partial_z}{(\bar{z}-\bar{z}_n)^{s-1}} + \frac{1}{r-1} \frac{L_{r-2,n} \partial_{\bar{z}}}{(z-z_n)^{r-1}} \right) \langle \mathcal{O}_{T\bar{T}}(z)X \rangle^{(0)}$$
$$+ \langle \mathcal{O}_{T\bar{T}}(z) dX \rangle^{(0)} \tag{44}$$

with divergences appearing only in the last term, removable by the local field renormalization (21) on $X$. The replacement $d \to d_R$ [from (24)] then follows.

The expression (44) is one of the main results of our work. A crucial difference between (20) and (44) is that in the latter, the operator $\mathcal{O}_{T\bar{T}}$ also flows under the deformation. That contribution can be isolated by subtracting $-\frac{1}{\pi} \int d^2 x' \langle \mathcal{O}_{T\bar{T}}(z') \mathcal{O}_{T\bar{T}}(z) X \rangle^{(0)}$, leading to

$$\langle \mathcal{O}_{T\bar{T}}^{(1)}(z) X \rangle^{(0)} = -\sum_n \sum_{r,s \geq 2} \left( \frac{1}{s-1} \frac{\bar{L}_{s-2,z} \partial_{z_n}}{(\bar{z}_n - \bar{z})^{s-1}} \right.$$
$$\left. + \frac{1}{r-1} \frac{L_{r-2,z} \partial_{\bar{z}_n}}{(z_n - z)^{r-1}} \right) \langle \mathcal{O}_{T\bar{T}}^{(0)}(z) X \rangle^{(0)}. \tag{45}$$

In the special case of a deformed free boson and $X = (\partial \phi \bar{\partial} \phi)^3(z_2)$, it can be shown that

$$\langle \mathcal{O}_{T\bar{T}}(z_1)(\partial \phi \bar{\partial} \phi)^3(z_2) \rangle^{(1)} = 0,$$
$$\langle \mathcal{O}_{T\bar{T}}^{(1)}(z_1)(\partial \phi \bar{\partial} \phi)^3(z_2) \rangle^{(0)} = \frac{9}{32\pi^3} \frac{1}{z_{12}^6 \bar{z}_{12}^6}, \tag{46}$$

consistent with standard perturbation theory (see Appendix B 2).

## IV. HIGHER-ORDER CORRECTIONS

### A. Correlators of undeformed operators

We now investigate the second-order correction in a correlator of undeformed operators on the Euclidean plane. While progress has been made on stress-tensor two-point functions at this order (see, for instance, [15]), calculations for more general correlators remain unexplored. Building on our first-order results, we use the relation (10) to rewrite the second-order correction as an integral involving the integrand in (44),

$$\langle X \rangle^{(2)} = -\frac{1}{2\pi} \int d^2 x \langle \mathcal{O}_{T\bar{T}}(z) X \rangle^{(1)}. \tag{47}$$

Due to the presence of functions with poles in the integration, we apply a point-splitting regularization for each stress tensor insertion and introduce an additional UV regulator $\varepsilon' \ll \varepsilon$. The integrand then becomes

$$\lim_{z \to z'} \partial_{\bar{z}}^{-1} \partial_z \langle (T(z+\varepsilon')\bar{T}(z))\bar{T}(z'+\varepsilon) X \rangle^{(0)}. \tag{48}$$

Details of the calculation are presented in Appendix C 2.

The final renormalized result for the second-order correction takes the form

$$\langle X_R \rangle^{(2)} = \frac{1}{2} \langle d_R^2 X \rangle^{(0)} + \sum_{m,n,i \text{ all distinct}} \sum_{r \geq 1}$$
$$\times \frac{1}{2} \left\langle \left[ \left( \frac{\log(\mu^2 |z_{in}|^2)}{z_{im}^r} + \sum_{p=1}^{r-1} \frac{1}{r-p} \frac{1}{z_{im}^{r-p} z_{nm}^p} + \frac{\log(|z_{im}|^2/|z_{in}|^2)}{z_{nm}^r} \right) L_{r-2,m} \bar{L}_{-1,n} \bar{L}_{-1,i} + \text{c.c.} \right] X \right\rangle^{(0)}$$
$$- \sum_{t>1} \frac{1}{2} \left\langle \left[ \frac{1}{t-1} \frac{1}{z_{nm}^r \bar{z}_{mi}^{t-1}} L_{r-2,m} \bar{L}_{-1,n} \bar{L}_{t-2,i} + \text{c.c.} \right] X \right\rangle^{(0)}$$
$$+ \sum_{s>1, t \geq 1} \frac{1}{2} \left\langle \left[ \binom{t+s-3}{s-2} \frac{1}{s-1} \frac{1}{\bar{z}_{in}^{s-1} \bar{z}_{ni}^{t-1}} \left( \frac{1}{z_{im}^r} - \frac{1}{z_{nm}^r} \right) L_{r-2,m} \bar{L}_{s-2,n} \bar{L}_{t-2,i} + \text{c.c.} \right] X \right\rangle^{(0)}, \tag{49}$$

where the renormalized operators $X_R = \prod_i \mathcal{O}_R(z_i)$ arise from a local renormalization given by

$$\mathcal{O}_R(z_i) = \mathcal{O}(z_i) + \frac{\lambda}{\pi} \int d^2x \left( \frac{[L_{-1}(z_i), [\bar{L}_{-1}(z_i), \mathcal{O}(z_i)]]}{(z - z_i + \mu)(\bar{z} - \bar{z}_i)} + \sum_{r,s \geq 1} \frac{[L_{r-2}(z_i), [\bar{L}_{s-2}(z_i), \mathcal{O}(z_i)]]}{(z - z_i + \varepsilon)^r (\bar{z} - \bar{z}_i)^s} \right)$$
$$+ \frac{\lambda^2}{2\pi} \int d^2x \sum_{r,t \geq 1} r \frac{\log(\mu^2 |z - z_i + \varepsilon'|^2)}{(z - z_i)^{r+1}(\bar{z} - \bar{z}_i + \varepsilon)^t} \mathrm{K}(L_{r-2}, L_{-1}, L_{t-2}; \mathcal{O})(z_i) + \text{c.c.}$$
$$+ \frac{\lambda^2}{2\pi} \int d^2x \sum_{r,t \geq 1, s > 1} \frac{r}{s-1} \frac{\mathrm{K}(L_{r-2}, L_{s-2}, L_{t-2}; \mathcal{O})(z_i)}{(z - z_i + \varepsilon')^{r+1}(\bar{z} - \bar{z}_i)^{s-1}(\bar{z} - \bar{z}_i + \varepsilon)^t} + \text{c.c.} + O(\lambda^3) \quad (50)$$

with

$$\mathrm{K}(L_{r-2}, L_{s-2}, L_{t-2}; \mathcal{O})(z_i) \equiv L_{r-2}(z_i)[\bar{L}_{s-2}(z_i), [\bar{L}_{t-2}(z_i), \mathcal{O}(z_i)]] + [L_{r-2}(z_i), \bar{L}_{s-2}(z_i)[\bar{L}_{t-2}(z_i), \mathcal{O}(z_i)]]$$
$$+ [L_{r-2}(z_i), [\bar{L}_{s-2}(z_i), \bar{L}_{t-2}(z_i)\mathcal{O}(z_i)]] - 2[L_{r-2}(z_i), [\bar{L}_{s-2}(z_i), [\bar{L}_{t-2}(z_i), \mathcal{O}(z_i)]]]. \quad (51)$$

In the expression for the kernel $\mathrm{K}(L_{r-2}, L_{s-2}, L_{t-2}; \mathcal{O})(z_i)$, note that in three of the terms at least one Virasoro generator $L$ appears as a multiplicative factor instead of acting on the field through a commutator. In this case, the factor $L_{r-2}(z_i)$ is "standalone" [i.e., not tied directly to $\mathcal{O}(z_i)$ through a commutator] and is free to act on all fields within the correlator. In effect, the combination

$$\frac{L_{r-2}(z_i)}{(z - z_i)^r}$$

is equivalent to the insertion of the stress tensor (T(z)) whose action on the correlator is to sum over contributions from all insertion points, $\sum_m L_{r-2,m} X$. These terms accounts for the divergences from the case when only two of the three indices $m$, $n$, $i$ are coincident. Since it automatically includes the contribution from the to-be-renormalized field (i.e., the triple-coincident case), we must subtract two duplicate instances to avoid overcounting.

For a two-point function of primary operators, (49) simplifies significantly, yielding

$$\langle \mathcal{O}_R(z) \mathcal{O}_R(w) \rangle^{(2)} = \frac{1}{2} (\log(\mu^2 |z - w|^2))^2 (\partial_z^2 \partial_{\bar{w}}^2 + \text{c.c.}) \langle \mathcal{O}(z) \mathcal{O}(w) \rangle^{(0)}$$
$$- \frac{1}{2} \log(\mu^2 |z - w|^2) \left( \frac{6h \partial_{\bar{w}}^2}{(z - w)^2} + \text{c.c.} \right) \langle \mathcal{O}(z) \mathcal{O}(w) \rangle^{(0)}$$
$$- \frac{1}{2} \left( \frac{4h\bar{h}(\bar{h} - 1)}{|z - w|^4} + \text{c.c.} \right) \langle \mathcal{O}(z) \mathcal{O}(w) \rangle^{(0)}. \quad (52)$$

Note that aside from the dominant leading $\log^2$ term (i.e., the term quadratic in log), the coefficients of the remaining terms depend on the renormalization scale $\mu$ and may be absorbed.

We now compare our result (52) to those in the literature. Nonperturbative two-point functions have been derived in three distinct frameworks: the resummation of leading logarithmic divergences at small momentum in [42], the JT-formalism analysis of large-momentum asymptotics in [43], and the world sheet computation of TsT-deformed vertex operators in [45] (with $w = 1$ corresponding to the double-trace $T\bar{T}$ deformation). While the results of [43,45] align precisely, they differ from [42] by a momentum-dependent factor.

This discrepancy is addressed in [43], where a perturbative first-order correction yields agreement in the leading logarithmic term (i.e., the term linear in log). The analysis therein emphasizes that, within perturbation theory, only the leading logarithmic term is scheme independent and thus appropriate for cross-comparison. Consistent with this observation, our findings indicate that the subleading contributions to the first- and second-order corrections depend on the renormalization scale $\mu$. Accordingly, a meaningful consistency check should focus on the leading logarithmic behavior.

To substantiate this, we expand the resummed two-point function from [42] [Eq. (4.28)] to second order in perturbation theory. The leading $\log^2$ term (quadratic in the logarithm) exactly matches the corresponding term in our result (52), thereby reinforcing the coherence of the leading logarithmic behavior across the various approaches.

### B. Stress tensor correlators

This section considers higher-order corrections to stress tensor correlators within the framework of the cutoff AdS/$T\bar{T}$-CFT duality [14,16]. Earlier works, such as [15,20], computed two-point and three-point correlators of the stress tensor up to the lowest nontrivial order. More recently, [31] extended these computations to two-loop order in Newton's constant $G$ for 3D gravity, providing a more comprehensive picture.

The analysis of stress tensor correlators is particularly important for testing cutoff AdS/$T\bar{T}$-CFT duality. These correlators exhibit consistent structures across conformal field theories, allowing for the derivation of universal conclusions. Moreover, the finite cutoff gravity dual for $T\bar{T}$-deformed CFTs is expected to hold in the pure gravity sector without additional matter fields. Hence, it is most appropriate to focus on pure stress tensor correlators within this duality proposal.

We begin by examining the second-order correction $\langle T(z)T(w)\rangle^{(2)}$ using the conservation equation method,

$$\langle T(z)T(w)\rangle^{(2)} = \partial_{\bar{z}}^{-1}\partial_z\partial_{\bar{w}}^{-1}\partial_w \langle \Theta(z)\Theta(w)\rangle^{(2)}$$
$$= \partial_{\bar{z}}^{-1}\partial_z\partial_{\bar{w}}^{-1}\partial_w \langle \mathcal{O}_{T\bar{T}}(z)\mathcal{O}_{T\bar{T}}(w)\rangle^{(0)}$$
$$= \frac{5c^2}{6}\frac{1}{(z-w)^6(\bar{z}-\bar{w})^2}, \quad (53)$$

where the holomorphic term from the antiderivative with respect to $\bar{z}$ is fixed to be zero by the spin constraint. This expression agrees with the direct perturbation theory results of [4].

Next, we extend the analysis to the third-order correction,

$$\langle \Theta(z)\Theta(w)\rangle^{(3)} = \langle \mathcal{O}_{T\bar{T}}(z)\mathcal{O}_{T\bar{T}}(w)\rangle^{(1)}$$
$$= \langle (T\bar{T} - (\Theta)^2)(z)(T\bar{T} - (\Theta)^2)(w)\rangle^{(1)}. \quad (54)$$

Here, various terms such as $\langle (\Theta)^2(z)(\Theta)^2(w)\rangle^{(1)}$ and $\langle (T\bar{T})(z)(\Theta)^2(w)\rangle^{(1)}$ on the rhs vanish upon using the trace relation. The remaining term $\langle (T\bar{T})(z)(T\bar{T})(w)\rangle^{(1)}$ can be extracted from the first-order correction to $\langle T(z_1)\bar{T}(z_2)T(z_3)\bar{T}(z_4)\rangle^{(1)}$ in (31) by taking the limits $z_1 \to z_2, z_3 \to z_4$, and then subtracting the divergent terms that arise from these limits

$$\langle \Theta(z)\Theta(0)\rangle^{(3)} = \langle (T\bar{T})(z)(T\bar{T})(0)\rangle^{(1)}$$
$$= \frac{c^2}{(z\bar{z})^5}(8\log(\mu^2|z|^2) - 8). \quad (55)$$

Applying the conservation equations leads to the third-order corrections for the other stress tensor two-point functions,

$$\langle T(z)\Theta(0)\rangle^{(3)} = \frac{c^2}{z^6\bar{z}^4}(10\log(\mu^2|z|^2) - 19/2), \quad (56)$$

$$\langle T(z)T(0)\rangle^{(3)} = \frac{c^2}{z^7\bar{z}^3}(20\log(\mu^2|z|^2) - 47/3). \quad (57)$$

Following similar steps, we compute the three-point function $\langle T(z_1)T(z_2)\Theta(z_3)\rangle^\lambda$ to its first nontrivial order. The result is

$$\langle T(z_1)T(z_2)\Theta(z_3)\rangle^{(2)} = \frac{c^2}{3\bar{z}_{13}^3}\left[\frac{1}{z_{23}^2}\left(\frac{1}{z_{12}^2 z_{13}^3} + \frac{1}{z_{12}^3 z_{13}^2}\right)\right]$$
$$+ (1 \leftrightarrow 2), \quad (58)$$

where invariance under the interchange $(1 \leftrightarrow 2)$ is imposed to fix the holomorphic terms from the conservation equations. As expected, all corrections preserve spin.

## V. CONCLUSIONS

In this study, we systematically computed various correlators in a general $T\bar{T}$-deformed CFT up to first and higher orders in $\lambda$. Standard perturbation theory proves inadequate for these calculations due to the absence of explicit expressions for the deformed operators, particularly the stress tensor in a general deformed CFT.

To address this difficulty, we developed a novel "conservation equation method." This approach utilizes the trace relation and conservation laws of the stress tensor to express higher-order corrections in terms of lower-order ones, at the cost of introducing certain indeterminate functions. By leveraging the correlator's structural symmetry and the unbroken spacetime symmetry of the deformed theory, we constrained and subsequently fixed these functions in our illustrative examples.

As an application, we computed first-order corrections to a variety of correlators, namely those with undeformed operators only, those with the stress tensor only, and those combining $\mathcal{O}_{T\bar{T}}$ and undeformed operators, i.e., $\langle O_{T\bar{T}}X\rangle^{(1)}$. From these results, we extracted the correlator involving the first-order deformation of $\mathcal{O}_{T\bar{T}}$, namely $\langle \mathcal{O}_{T\bar{T}}^{(1)}X\rangle^{(0)}$, which is inaccessible in standard perturbation theory. We also obtained an expression for the second-order correction to correlators of undeformed operators, $\langle X \rangle^{(2)}$, based on the result for $\langle \mathcal{O}_{T\bar{T}} X \rangle^{(1)}$. Higher-order corrections to several pure stress tensor correlators are likewise examined. Our results demonstrate consistency with examples computed in the deformed free boson CFT, as well as with previous works in both perturbative and nonperturbative regimes.

We find that undeformed operators' correlators exhibit both logarithmic and power UV divergences at first order, and a double logarithmic divergence at second order, accompanied by single-logarithm and power divergences. All divergences in the bare results can be eliminated by local field renormalizations, which we explicitly performed for general undeformed operators at the first and second orders.

The techniques used for calculating first-order corrections to stress tensor correlators and the mixed correlator $\langle \mathcal{O}_{T\bar{T}} X \rangle^{\lambda}$ allow for calculations to arbitrarily high orders in principle, though challenges may arise when addressing undetermined holomorphic terms from conservation equations. Whether these terms can be handled consistently in all meaningful correlators remains an open question.

A promising avenue for future investigation is extending this approach to deformed finite-size and/or finite-temperature CFTs. In these scenarios, the trace relation no longer holds, and the conservation equations fail to capture essential information such as one-point functions. Alternative principles, such as modular covariance, could potentially offer constraints on correlators in these contexts. Furthermore, a noteworthy prospect lies in validating the cutoff-AdS/$T\bar{T}$-CFT duality by computing correlators in a deformed CFT on a torus and comparing them with results from the gravity side [21,52], obtained using the prescription in [14,16]. Insights from the gravitational perspective could lead to the formulation of a generalized trace relation.

Another longstanding challenge is to develop a robust prescription for computing holographic mixed correlators as well as correlators of nonflowing fields (where "nonflowing fields" designate the matter fields in the bulk). It has been proposed that when matter fields are present the deformed CFT is dual to a gravitational theory with mixed boundary conditions [23]. Reproducing a general expression for the first-order correction to the mixed correlator would provide compelling evidence in support of the holographic proposal.

## ACKNOWLEDGMENTS

We thank John Cardy for valuable advice and insightful conversations. This work has received support from the National Natural Science Foundation of China through Grants No. 12475053, No. 12235016, No. 12075101, No. 12047569, No. 12235016, and No. 12105113. S. H. acknowledges financial assistance from the Fundamental Research Funds for the Central Universities and the Max Planck Partner Group.

## APPENDIX A: PROOF OF THE VANISHING OF THE DISCONNECTED TERM

We show that the disconnected term $\langle \mathcal{O}_{T\bar{T}}(z) \rangle^{\lambda} \langle (\prod_i O_i^{\lambda}(z_i)) \rangle^{\lambda}$ in the flow equation (2) vanishes. Since this term is proportional to $\langle \mathcal{O}_{T\bar{T}}(z) \rangle^{\lambda}$, it suffices to show that this one-point function vanishes. In early works [1], it was shown that on a Euclidean plane, the relation $\langle \mathcal{O}_{T\bar{T}} \rangle^{\lambda} = (\langle \Theta \rangle^{\lambda})^2$ holds. Expanding both sides in powers of $\lambda$, we obtain

$$\langle \mathcal{O}_{T\bar{T}} \rangle^{(n)} = \sum_{i+j=n} \langle \Theta \rangle^{(i)} \langle \Theta \rangle^{(j)}. \quad (A1)$$

Using the trace relation gives

$$\langle \Theta \rangle^{(n+1)} = \langle \mathcal{O}_{T\bar{T}} \rangle^{(n)} = \sum_{i+j=n} \langle \Theta \rangle^{(i)} \langle \Theta \rangle^{(j)}, \quad (A2)$$

which indicates higher-order corrections to $\langle \Theta \rangle^{\lambda}$ can always be expressed as a sum of products of its lower-order corrections. Conformal invariance of the undeformed theory implies the vanishing of $\langle \Theta \rangle^{(0)}$. Therefore, by induction, we conclude that $\langle \Theta \rangle^{\lambda}$ vanishes to all orders. This, in turn, implies that $\langle \mathcal{O}_{T\bar{T}} \rangle^{\lambda}$ vanishes as well.

## APPENDIX B: CONSISTENCY CHECKS AND DETAILS

### 1. First-order correction to the stress tensor correlator

To verify the result in Sec. III B 2, we compute $\langle T(z_1) \bar{T}(z_2) T(z_3) \bar{T}(z_4) \rangle^{(1)}$ using standard perturbation theory. While this method is typically impractical for generic deformed CFTs—owing to the lack of a prescription for the deformed stress tensor—it becomes feasible for certain seeds, such as free theories, where explicit forms are known. In the following, we focus on a deformed massless free boson, with

$$T^{(0)} = -2\pi(\partial\phi)^2, \qquad \mathcal{L}^{(1)} = \frac{1}{\pi} T^{(0)} \bar{T}^{(0)},$$
$$T^{(1)} = -8\pi^2 (\partial\phi)^3 \bar{\partial}\phi, \qquad \mathcal{O}_{T\bar{T}}^{(1)} = 32\pi^3 (\partial\phi \bar{\partial}\phi)^3. \quad (B1)$$

In this example, contributions from the flow of action, such as

$$\langle T^{(1)}(z_1) \bar{T}^{(0)}(z_2) T^{(0)}(z_3) \bar{T}^{(0)}(z_4) \rangle^{(0)} + (1 \leftrightarrow 3),$$

vanish upon Wick contraction. Therefore, we have

$$\langle T(z_1)\bar{T}(z_2)T(z_3)\bar{T}(z_4)\rangle^{(1)} = -\frac{1}{\pi}\int d^2z \langle \mathcal{O}_{T\bar{T}}(z)T(z_1)\bar{T}(z_2)T(z_3)\bar{T}(z_4)\rangle^{(0)}$$

$$= -\frac{1}{\pi}\int d^2z \left(\frac{\partial_{z_1}}{(z-z_1)} + \frac{2}{(z-z_1)^2} + (1\leftrightarrow 3)\right)\left(\frac{\partial_{\bar{z}_4}}{(\bar{z}-\bar{z}_4)} + \frac{2}{(\bar{z}-\bar{z}_4)^2} + (4\leftrightarrow 2)\right)$$

$$\times \langle T(z_1)\bar{T}(z_2)T(z_3)\bar{T}(z_4)\rangle^{(0)}$$

$$= \frac{c^2}{z_{13}^4 \bar{z}_{24}^4}\left[\left(-\frac{4\log(|z_{14}|^2/\epsilon^2)}{z_{13}\bar{z}_{24}} + \frac{2}{z_{13}\bar{z}_{14}} + \frac{2}{z_{23}\bar{z}_{24}}\right) + (1\leftrightarrow 3) + (4\leftrightarrow 2) + (1\leftrightarrow 3, 4\leftrightarrow 2)\right].$$

(B2)

Renormalizing the stress tensors as in (21) would replace the divergence $1/\epsilon$ by $\mu$. We thus find perfect agreement with the previous calculation (31).

### 2. First-order correction to mixed correlators

In Sec. III C, a general expression (44) for the first-order correction to mixed correlators and the stress tensor flow contribution are derived. As an illustrative case, consider $X = (\partial\phi\bar{\partial}\phi)^3$.

Using (44), the term involving the logarithm in the first line vanishes because the undeformed correlator itself vanishes. The remaining terms are

$$\langle \mathcal{O}_{T\bar{T}}(z)L_n\mathcal{O}_{T\bar{T}}^{(1)}(z_2)\rangle^{(0)} = \frac{1}{2\pi i}\oint_{z_2} dz'(z'-z_2)^{n+1}\langle \mathcal{O}_{T\bar{T}}(z)T(z')\mathcal{O}_{T\bar{T}}^{(1)}(z_2)\rangle^{(0)}$$

$$= -\frac{1}{2\pi i}\oint_{z} dz'(z'-z_2)^{n+1}\langle \mathcal{O}_{T\bar{T}}(z)T(z')\mathcal{O}_{T\bar{T}}^{(1)}(z_2)\rangle^{(0)} \quad \text{(Reversing the contour)}$$

$$= -\frac{1}{2\pi i}\oint_{z} dz'(z'-z_2)^{n+1}\left(\frac{\partial_z}{z'-z} + \frac{2}{(z'-z)^2}\right)\langle \mathcal{O}_{T\bar{T}}(z)\mathcal{O}_{T\bar{T}}^{(1)}(z_2)\rangle^{(0)}$$

$$-\frac{1}{2\pi i}\oint_{z} dz'(z'-z_2)^{n+1}\frac{c/2}{(z'-z)^4}\langle \bar{T}(z)\mathcal{O}_{T\bar{T}}^{(1)}(z_2)\rangle^{(0)} = 0,$$

(B3)

where in the first line, we used the definition of $L_n$, and in the last equality we used the fact that the correlators $\langle \mathcal{O}_{T\bar{T}}(z)\mathcal{O}_{T\bar{T}}^{(1)}\rangle^{(0)}$ and $\langle \bar{T}(z)\mathcal{O}_{T\bar{T}}^{(1)}(z_2)\rangle^{(0)}$ vanish upon Wick contraction.

We now reproduce the result with standard perturbation theory. The contribution from the flow of $\mathcal{O}_{T\bar{T}}(z_1)$, i.e., the first term on the rhs of (3) is

$$\langle \mathcal{O}_{T\bar{T}}^{(1)}(z_1)(\partial\phi\bar{\partial}\phi)^3(z_2)\rangle^{(0)} = 32\pi^3\langle(\partial\phi\bar{\partial}\phi)^3(z_1)(\partial\phi\bar{\partial}\phi)^3(z_2)\rangle^{(0)} = \frac{9}{32\pi^3}\frac{1}{z_{12}^6\bar{z}_{12}^6}.$$

(B4)

The second term on the rhs of (3), which is the contribution from the flow of action, is given by

$$-\frac{1}{\pi}\int d^2x \langle \mathcal{O}_{T\bar{T}}(z)\mathcal{O}_{T\bar{T}}(z_1)(\partial\phi\bar{\partial}\phi)^3(z_2)\rangle^{(0)} = -\frac{1}{\pi}\int d^2x \langle (2\pi)^2(\partial\phi\bar{\partial}\phi)^2(z)(2\pi)^2(\partial\phi\bar{\partial}\phi)^2(z_1)(\partial\phi\bar{\partial}\phi)^3(z_2)\rangle^{(0)},$$

(B5)

where the only contribution comes from the term involving the contact term contraction $\langle \bar{\partial}\phi(z)\partial\phi(z_1)\rangle = \frac{1}{4\pi}\bar{\partial}\frac{1}{z-z_1} = \frac{1}{4\pi}\partial\frac{1}{\bar{z}-\bar{z}_1} = \frac{1}{4}\delta^2(x-x_1)$. All other contractions, such as $\langle \bar{\partial}\phi(z)\partial\phi(z_3)\rangle$, lead to zero. Thus, (B5) becomes

$$-\frac{1}{\pi}(2\pi)^4 \cdot 2 \cdot 2^2 \int d^2x \frac{1}{4}\delta^2(x-x_1)\langle((\partial\phi)^2\bar{\partial}\phi)(z)(\partial\phi(\bar{\partial}\phi)^2)(z_1)(\partial\phi\bar{\partial}\phi)^3(z_2)\rangle^{(0)}$$

$$= -32\pi^3\langle(\partial\phi\bar{\partial}\phi)^3(z_1)(\partial\phi\bar{\partial}\phi)^3(z_2)\rangle^{(0)}.$$

(B6)

Summing (B4) and (B5) up leads to

$$\langle \mathcal{O}_{T\bar{T}}(z_1)(\partial\phi\bar{\partial}\phi)^3(z_2)\rangle^{(1)} = 0. \quad (B7)$$

Namely, the contribution from the flow of action and the flow of stress tensor cancel out.

As another example, we compute $\langle \mathcal{O}_{T\bar{T}}^{(1)}(z_1)(\partial\phi\bar{\partial}\phi)^3(z_2)\rangle^{(0)}$, i.e., the contribution from the flow of action, using Eq. (45). For the deformed massless free boson, the undeformed $T\bar{T}$ operator is given by $\mathcal{O}_{T\bar{T}}^{(0)}(z_1) = (2\pi)^2(\partial\phi\bar{\partial}\phi)^2(z_1)$. We can write $(\bar{\partial}\phi)^3(z_2)$ as

$$\lim_{5\to 2, 4\to 2, 3\to 2} \partial_{\bar{z}_3}\partial_{\bar{z}_4}\partial_{\bar{z}_5}[\phi(z_3)\phi(z_4)\phi(z_5)],$$

then evaluate $\langle \mathcal{O}_{T\bar{T}}^{(1)}(z_1)(\partial\phi)^3(z_2)[\phi(z_3)\phi(z_4)\phi(z_5)]\rangle^{(0)}$, and take the derivatives and limits at the end of the calculation.[6] Using Eq. (45), we find the relevant terms to be

$$\langle \mathcal{O}_{T\bar{T}}^{(1)}(z_1)(\partial\phi)^3(z_2)[\phi(z_3)\phi(z_4)\phi(z_5)]\rangle^{(0)}$$
$$= (2\pi)^2 \sum_{n=3,4,5} \left\langle \frac{\bar{L}_{s-2,1}\partial_{z_n}}{\bar{z}_{n1}^{s-1}}(\partial\phi\bar{\partial}\phi)^2(z_1)(\partial\phi)^3(z_2)[\phi(z_3)\phi(z_4)\phi(z_5)] \right\rangle^{(0)}. \quad (B8)$$

$(\partial\phi)^3(z_2)$ can be rewritten as $(\partial\phi)^2(z_2)\partial\phi(z_2) = -\frac{1}{2\pi}T(z_2)\partial\phi(z_2)$, allowing us to use the conformal Ward identity to evaluate this term. The contributing terms are the connected part acting on $(\partial\phi\bar{\partial}\phi)^2(z_1)$

$$\langle \mathcal{O}_{T\bar{T}}^{(1)}(z_1)(\partial\phi)^3(z_2)[\phi(z_3)\phi(z_4)\phi(z_5)]\rangle^{(0)}$$
$$= (-2\pi) \sum_{n=3,4,5} \left\langle \frac{\bar{L}_{s-2,1}\partial_{z_n}}{\bar{z}_{n1}^{s-1}}\left(\frac{\partial_{z_1}}{z_{21}} + \frac{2}{z_{21}^2}\right)(\partial\phi\bar{\partial}\phi)^2(z_1)\partial\phi(z_2)[\phi(z_3)\phi(z_4)\phi(z_5)] \right\rangle^{(0)}. \quad (B9)$$

The above is nonzero only when $s = 2$ or $s = 4$; in the latter case, $(\bar{\partial}\phi)^2(z_1)$ is annihilated, and the entire correlator vanishes upon Wick contraction. The surviving terms are

$$\langle \mathcal{O}_{T\bar{T}}^{(1)}(z_1)(\partial\phi)^3(z_2)[\phi(z_3)\phi(z_4)\phi(z_5)]\rangle^{(0)}$$
$$= (-2\pi) \sum_{n=3,4,5} \left\langle \frac{2\partial_{z_n}}{\bar{z}_{n1}}\left(\frac{\partial_{z_1}}{z_{21}} + \frac{2}{z_{21}^2}\right)(\partial\phi\bar{\partial}\phi)^2(z_1)\partial\phi(z_2)[\phi(z_3)\phi(z_4)\phi(z_5)] \right\rangle^{(0)}. \quad (B10)$$

Performing Wick contractions gives

$$\langle \mathcal{O}_{T\bar{T}}^{(1)}(z_1)(\partial\phi)^3(z_2)[\phi(z_3)\phi(z_4)\phi(z_5)]\rangle^{(0)}$$
$$= 4(-2\pi)\frac{1}{(4\pi)^4}\frac{2}{\bar{z}_{31}}\left(\frac{\partial_{z_1}}{z_{21}} + \frac{2}{z_{21}^2}\right)\frac{1}{z_{12}^2}\left(\frac{1}{z_{13}\bar{z}_{14}\bar{z}_{15}} + \text{perm.}(3,4,5)\right). \quad (B11)$$

Taking the $\partial_{\bar{z}_3}\partial_{\bar{z}_4}\partial_{\bar{z}_5}$ derivatives and the $z_3 \to z_2, z_4 \to z_2, z_5 \to z_2$ limits results in

$$\langle \mathcal{O}_{T\bar{T}}^{(1)}(z_1)(\partial\phi\bar{\partial}\phi)^3(z_2)\rangle^{(0)} = \frac{9}{32\pi^3}\frac{1}{z_{12}^6 \bar{z}_{12}^6}, \quad (B12)$$

which agrees with the result obtained using contractions and the explicit form of $\mathcal{O}_{T\bar{T}}^{(1)}(z_1)$

$$\langle \mathcal{O}_{T\bar{T}}^{(1)}(z_1)(\partial\phi\bar{\partial}\phi)^3(z_2)\rangle^{(0)} = 32\pi^3\langle(\partial\phi\bar{\partial}\phi)^3(z_1)(\partial\phi\bar{\partial}\phi)^3(z_2)\rangle^{(0)} = \frac{9}{32\pi^3}\frac{1}{z_{12}^6 \bar{z}_{12}^6}. \quad (B13)$$

---

[6]$\lim_{5\to 2,4\to 2,3\to 2}\bar{\partial}\phi(z_3)\bar{\partial}\phi(z_4)\bar{\partial}\phi(z_5)$ is nothing but the definition of the product of fields $(\bar{\partial}\phi)^3(z_2)$; Here the only trick is taking the derivatives at the end of the calculation, which is justified by the rule $\partial_{z_m}\langle \mathcal{O}_{T\bar{T}}^{(1)}(z)X\rangle^{(0)} = \langle \mathcal{O}_{T\bar{T}}^{(1)}(z)(\partial_{z_m}X)\rangle^{(0)}$. Further explanation regarding this rule is provided at the end of this section.

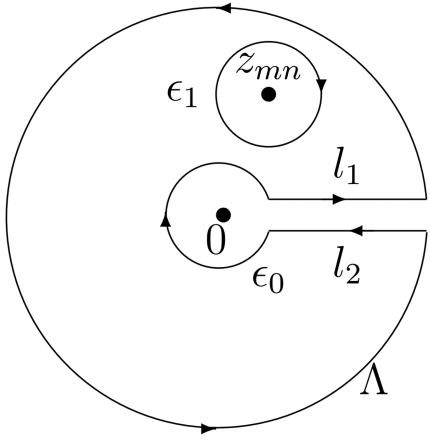

FIG. 1. Contour for the $r=1$, $s=1$ integral.

In the computation of these examples, we have utilized the relation $\langle \partial_{z_m} X \rangle^\lambda = \partial_{z_m} \langle X \rangle^\lambda$ predicated on the translational invariance of the vacuum state. Discrepancies that seem to violate this rule [see (20) and (45)] arise primarily from the integral regularization process, which can obscure certain cancellations. In computations that bypass such regularization, these irregularities do not appear. For the deformed massless free boson, we treat $\phi$ and its correlators as the basic objects, from which correlators of $\partial\phi$ and $\partial\bar\phi$ are derived.

## APPENDIX C: USEFUL INTEGRALS

In this appendix we compute some integrals used in constructing the expressions for the first and second-order correction to a correlator of undeformed operators. Divergent integrals will be regularized by cutting out a tiny disk of radius $\varepsilon \to 0$ around the poles.

### 1. Integrals for the first-order correction

We will perform the integrals needed in computing first-order correction (19), which in general take the form

$$I_{m\bar n}^{r\bar s} = \int \frac{d^2 x}{(z-z_m)^r (\bar z - \bar z_n)^s} = \frac{i}{2} \int \frac{d^2 z}{(z-z_m)^r (\bar z - \bar z_n)^s}. \quad (C1)$$

For the case of $r=1$, $s=1$

$$I_{m\bar n}^{1\bar 1} = \frac{i}{2}\int \frac{d^2z}{(z-z_m)(\bar z - \bar z_n)} = \frac{i}{2}\int \frac{d^2 z}{(z-z_{mn})\bar z} = \frac{i}{2}\int d^2 z \partial_{\bar z}\left(\frac{\log \bar z}{(z-z_{mn})\bar z}\right) = -\frac{i}{2}\oint dz \frac{\log \bar z}{z-z_{mn}}. \quad (C2)$$

The integrand has a branch cut from $z=0$ to $\infty$ and a pole $z=z_{mn}$, see Fig. 1

$$-\frac{i}{2}\oint dz\frac{\log \bar z}{z-z_{mn}} = I_\Lambda + I_{l_1+l_2} + I_{\epsilon_0} + I_{\epsilon_1}, \quad (C3)$$

$$I_\Lambda = \frac{1}{2}\int_0^{2\pi} \Lambda e^{i\theta} d\theta \frac{\log\Lambda - i\theta}{\Lambda e^{i\theta}-z_{mn}} = \frac{1}{2}\int_0^{2\pi} d\theta(\log\Lambda - i\theta) + \frac{z_{mn}}{2}\int_0^{2\pi}d\theta \frac{\log\Lambda - i\theta}{\Lambda e^{i\theta}-z_{mn}}$$
$$= \pi \log \Lambda - \pi^2 i + 0;$$

$$I_{\epsilon_0} = -\frac{1}{2}\int_0^{2\pi}\epsilon e^{i\theta}d\theta \frac{\log\epsilon - i\theta}{\epsilon e^{i\theta}-z_{mn}} \to 0;$$

$$I_{\epsilon_1} = \frac{i}{2}\oint_{|z-z_{mn}|=\epsilon} dz \frac{\log\left(\bar z_{mn}+\frac{\epsilon^2}{z-z_{mn}}\right)}{z - z_{mn}} = \frac{i}{2}\oint_{|z'|=\epsilon}\frac{dz'}{z'}\log\left(\bar z_{mn}+\frac{\epsilon^2}{z'}\right)$$
$$= \frac{i}{2}\oint_{|z'|=\epsilon_1}\frac{dz'}{z'}\left[\log \bar z_{mn} + \log\left(1+\frac{\epsilon_1^2}{z'\bar z_{mn}}\right)\right] = -\pi \log \bar z_{mn} - \frac{i}{2}\sum_{n=1}^\infty \frac{1}{n}\oint \frac{dz}{z^{n+1}}\left(\frac{\epsilon_1^2}{\bar z_{nm}}\right)^n$$
$$= -\pi \log \bar z_{mn} - 0;$$

$$I_{l_1+l_2} = -\frac{i}{2}\int_\epsilon^\Lambda dx \frac{\log x}{x-z_{mn}} + \frac{i}{2}\int_\epsilon^\Lambda dx \frac{\log x - 2\pi i}{x - z_{mn}} = \pi\int_\epsilon^\Lambda \frac{dx}{x-z_{mn}} = \pi \log \Lambda - \pi \log z_{nm}.$$

Summing up

$$I_{m\bar n}^{1\bar 1} = 2\pi \log \Lambda - \pi \log(-|z_{mn}|^2) - \pi^2 i = -\pi \log(|z_{mn}|^2/\Lambda^2), \quad (C4)$$

where $\Lambda$ is an IR cutoff. For $r = 1$, $s \geq 1$

$$\begin{aligned}
I^{1\bar{s}}_{m\bar{n}} &= \frac{i}{2}\int \frac{d^2z}{(z-z_{mn})\bar{z}^s} = \frac{i}{2}\int_{|z|>|z_{mn}|} \frac{d^2z}{\left(1-\frac{z_{mn}}{z}\right)z\bar{z}^s} - \frac{i}{2}\int_{|z|<|z_{mn}|} \frac{d^2z}{\left(1-\frac{z}{z_{mn}}\right)z_{mn}\bar{z}^s} \\
&= \frac{i}{2}\int_{|z|>|z_{mn}|} \frac{d^2z}{z\bar{z}^s}\sum_{i=0}^{\infty}\left(\frac{z_{mn}}{z}\right)^i - \int_{|z|<|z_{mn}|} \frac{d^2z}{z_{mn}\bar{z}^s}\sum_{i=0}^{\infty}\left(\frac{z}{z_{mn}}\right)^i \\
&= \frac{i}{2}\int_{|z|>|z_{mn}|} d^2z\frac{z^{s-1}_{mn}}{z^s\bar{z}^s} - 0 = \int_0^{2\pi}d\theta\int_{|z_{mn}|}^{\infty}\rho d\rho\frac{z^{s-1}_{mn}}{\rho^{2s}} = \frac{\pi}{1-s}\frac{z^{s-1}_{mn}}{\rho^{2s-2}}\bigg|_{\rho=|z_{mn}|}^{\infty} = \frac{\pi}{s-1}\frac{1}{\bar{z}^{s-1}_{mn}}.
\end{aligned}$$ (C5)

Integrals with $r \geq 2$, $s \geq 2$ vanish because they are $\partial^{r-1}_{z_n}\partial^{s-1}_{\bar{z}_m}$ derivatives of the $r = 1$, $s = 1$ integral. We may also evaluate these integrals by using Stokes's theorem

$$\begin{aligned}
I^{1\bar{s}}_{m\bar{n}} &= \frac{i}{2}\int \frac{d^2z}{(z-z_{mn})\bar{z}^s} = -\frac{i}{2}\frac{1}{s-1}\int d^2z\partial_{\bar{z}}\left(\frac{1}{(z-z_{mn})\bar{z}^{s-1}}\right) \\
&= \frac{i}{2}\frac{1}{s-1}\left[\oint_{|z|=\Lambda}\frac{dz}{(z-z_{mn})\bar{z}^{s-1}} - \oint_{|z-z_{mn}|=\epsilon}\frac{dz}{(z-z_{mn})\bar{z}^{s-1}}\right] \\
&= 0 - \frac{i}{2}\frac{1}{s-1}\oint_{|z-z_{mn}|=\epsilon}\frac{dz}{(z-z_{mn})\left(\frac{\epsilon^2}{z-z_{mn}}+\bar{z}\right)^{s-1}} \\
&= -\frac{i}{2}\frac{2\pi i}{s-1}\frac{1}{\bar{z}^{s-1}_{mn}} + \mathcal{O}(\epsilon) = \frac{\pi}{s-1}\frac{1}{\bar{z}^{s-1}_{mn}}.
\end{aligned}$$ (C6)

Collecting all the integrals above gives

$$\begin{aligned}
\langle X\rangle^{(1)} &= \sum_{m\neq n}\left\langle \left(\log(|z_{mn}|^2/\Lambda^2)\partial_{z_m}\partial_{\bar{z}_n} - \sum_{s\geq 2}\frac{1}{s-1}\frac{\bar{L}_{s-2,n}\partial_{z_m}}{\bar{z}^{s-1}_{mn}} - \sum_{r\geq 2}\frac{1}{r-1}\frac{L_{r-2,m}\partial_{\bar{z}_n}}{z^{r-1}_{nm}}\right)X\right\rangle^{(0)} \\
&+ \sum_{m=n}\left\langle \left(\log(\epsilon^2/\Lambda^2)\partial_{z_m}\partial_{\bar{z}_n} - \sum_{s\geq 2}\frac{1}{s-1}\frac{\bar{L}_{s-2,n}\partial_{z_m}}{\epsilon^{s-1}} - \sum_{r\geq 2}\frac{1}{r-1}\frac{L_{r-2,m}\partial_{\bar{z}_n}}{\epsilon^{r-1}}\right)X\right\rangle^{(0)}.
\end{aligned}$$ (C7)

The dependence on the IR cutoff cancels upon summation over the indices $m$, $n$, as can be seen by rewriting $\sum_{m=n}\partial_{z_m}\partial_{\bar{z}_n}$ as $\sum_{m\neq n}\partial_{z_m}\partial_{\bar{z}_n}$ by translational invariance of correlators. (C7) then becomes (20).

### 2. Integrals for the second-order correction

To obtain the second-order correction, we need to evaluate the integral on the rhs of (47), which amounts to evaluating

$$\begin{cases} \left(-\frac{1}{2}\right)\frac{1}{\pi}\int d^2x\left\langle\left(\frac{r}{s-1}\frac{L_{r-2,m}\bar{L}_{s-2,n}\bar{L}_{t-2,i}}{(z-z_m)^{r+1}(\bar{z}-\bar{z}_n)^{s-1}(\bar{z}-\bar{z}_i)^t} + \text{c.c.}\right)X\right\rangle^{(0)}, & \text{for } s > 1, \\ \left(-\frac{1}{2}\right)\frac{1}{\pi}\int d^2x\left\langle\left(r\frac{L_{r-2,m}\bar{L}_{-1,n}\bar{L}_{t-2,i}}{(z-z_m)^{r+1}(\bar{z}-\bar{z}_i)^t}\log(\mu^2|z-z_n|^2) + \text{c.c.}\right)X\right\rangle^{(0)}, & \text{for } s = 1. \end{cases}$$ (C8)

In the following computation, we assume the coordinates $z_m$, $z_n$, $z_i$ do not coincide with each other. The cases where two or more of $z_m$, $z_n$, $z_i$ coincide can be obtained by taking coincidence limits.

Let us focus on the integral in the second line of (C8) with $s = t = 1$. Decomposing the logarithm as $\log(\mu^2|z-z_n|^2) = \log(\mu(z-z_n))$ produces two terms, the first one being

$$\frac{1}{\pi}\int d^2x \frac{\log(\mu(z-z_n))}{(z-z_m)^{r+1}(\bar{z}-\bar{z}_i)} = \frac{1}{\pi(-r)}\frac{i}{2}\int d^2z \partial_{\bar{z}}\left(\frac{\log(\mu(z-z_n))}{(z-z_m)^r(\bar{z}-\bar{z}_i)}\right) - \frac{1}{\pi(-r)}\frac{i}{2}\int d^2z \frac{\partial_z \log(\mu(z-z_n))}{(z-z_m)^r(\bar{z}-\bar{z}_i)}$$

$$= -\frac{1}{2\pi i r}\left(-\oint_{|z-z_i|=\epsilon} - \oint_{|z-z_n|=\epsilon} - \oint_{|z-z_m|=\epsilon} + \oint_{l_n} + \oint_{|z|=\Lambda}\right) d\bar{z} \frac{\log(\mu(z-z_n))}{(z-z_m)^r(\bar{z}-\bar{z}_i)}$$

$$+ \frac{1}{\pi r}\frac{i}{2}\int d^2z \frac{1}{(z-z_m)^r(z-z_n)(\bar{z}-\bar{z}_i)}, \tag{C9}$$

where $\oint_{l_n}$ denotes the contour integral along the branch cut of $\log(\mu(z-z_n))$. The integral around $z_n$ is

$$\oint_{|z-z_n|=\epsilon} d\bar{z}\log(\mu(z-z_n)) = \lim_{\epsilon\to 0}\int_0^{2\pi}d\theta \epsilon e^{-i\theta}\log(\mu\epsilon e^{i\theta}) = 0. \tag{C10}$$

Similarly, the integral around $z_m$ vanishes: $\oint_{|z-z_m|=\epsilon} d\bar{z}\frac{1}{(z-z_m)^r} = 0$. The integral around $z_i$ is

$$\frac{1}{2\pi i r}\oint_{|z-z_i|=\epsilon} d\bar{z}\frac{\log(\mu(z-z_n))}{(z-z_m)^r(\bar{z}-\bar{z}_i)} = -\frac{1}{r}\frac{\log(\mu z_{in})}{z_{im}^r}. \tag{C11}$$

The integral along the branch cut of $\log(\mu(z-z_n))$ is

$$-\frac{1}{2\pi i r}\oint_{l_n} d\bar{z}\frac{\log(\mu(z-z_n))}{(z-z_m)^r(\bar{z}-\bar{z}_i)} = \frac{1}{r}\int_{x_n}^{\Lambda}dx\frac{1}{(z-z_m)^r(\bar{z}-\bar{z}_i)}\bigg|_{y=y_n}, \tag{C12}$$

which can be canceled by (C17) as we will see in a moment. Finally, the integral over a circle at infinity $\Lambda\to\infty$ is

$$-\frac{1}{2\pi i r}\oint_{|z|=\Lambda} dz\frac{\log(\mu(z-z_n))}{(z-z_m)^r(\bar{z}-\bar{z}_i)} \sim -\frac{1}{2\pi i r}\int_0^{2\pi}\Lambda e^{i\theta}d\theta \frac{\log(\mu\Lambda)+i\theta}{(\Lambda e^{i\theta})^r\Lambda e^{-i\theta}}, \tag{C13}$$

which goes to zero in the limit $\Lambda\to\infty$.

Next, consider the three-pole integral in the last line of (C9). The decomposition rule $\frac{1}{(z-z_m)(z-z_n)} = \frac{1}{z_{mn}}\left(\frac{1}{z-z_m} - \frac{1}{z-z_n}\right)$ can be repeatedly applied to this integral until it is written in terms of two-pole integrals $I_{m\bar{n}}^{r\bar{s}}$, resulting in

$$\frac{i}{2\pi r}\int d^2z\frac{1}{(z-z_m)^r(z-z_n)(\bar{z}-\bar{z}_i)} = \frac{i}{2\pi r}\frac{1}{z_{nm}^{r-1}}\int d^2z\frac{1}{(z-z_m)(z-z_n)(\bar{z}-\bar{z}_i)}$$

$$- \frac{i}{2\pi r}\sum_{p=1}^{r}\frac{1}{z_{nm}^p}\int d^2z\frac{1}{(z-z_m)^{r-p+1}(\bar{z}-\bar{z}_i)}$$

$$= -\frac{1}{r}\left(\frac{1}{z_{nm}^r}\log\frac{|z_{im}|^2}{|z_{in}|^2} + \sum_{p=1}^{r-1}\frac{1}{r-p}\frac{1}{z_{im}^{r-p}z_{nm}^p}\right). \tag{C14}$$

We now proceed to the terms containing $\log(\mu(\bar{z}-\bar{z}_n))$

$$\frac{1}{\pi}\int d^2x \frac{\log(\mu(\bar{z}-\bar{z}_n))}{(z-z_m)^{r+1}(\bar{z}-\bar{z}_i)} = \frac{1}{\pi(-r)}\frac{i}{2}\int d^2z \partial_{\bar{z}}\left(\frac{\log(\mu(\bar{z}-\bar{z}_n))}{(z-z_m)^r(\bar{z}-\bar{z}_i)}\right)$$

$$= -\frac{1}{(2\pi i)r}\left(-\oint_{|z-z_n|=\epsilon} - \oint_{|z-z_i|=\epsilon} + \oint_{l_n} + \oint_{|z|=\Lambda}\right)d\bar{z}\frac{\log(\mu(\bar{z}-\bar{z}_n))}{(z-z_m)^r(\bar{z}-\bar{z}_i)}. \tag{C15}$$

We find the nonvanishing terms to be

$$\frac{1}{(2\pi i)r}\oint_{|z-z_i|=\epsilon} d\bar{z}\frac{\log(\mu(\bar{z}-\bar{z}_n))}{(z-z_m)^r(\bar{z}-\bar{z}_i)} = -\frac{1}{r}\frac{\log(\mu\bar{z}_{in})}{z_{im}^r}, \tag{C16}$$

and

$$-\frac{1}{(2\pi i)r}\oint_{l_n} d\bar{z}\frac{\log(\mu(\bar{z}-\bar{z}_n))}{(z-z_m)^r(\bar{z}-\bar{z}_i)}$$
$$=-\frac{1}{r}\int_{x_n}^{\Lambda} dx\frac{1}{(z-z_m)^r(\bar{z}-\bar{z}_i)}\bigg|_{y=y_n}. \quad (C17)$$

The latter cancels out the term in (C12).[7] Collecting all the results, we find the $s=1$, $t=1$ integral to be

$$\frac{1}{\pi}\int d^2x\frac{\log(\mu^2|z-z_n|^2)}{(z-z_m)^{r+1}(\bar{z}-\bar{z}_i)}$$
$$=-\frac{1}{r}\left(\frac{\log(\mu^2|z_{\rm in}|^2)}{z_{im}^r}+\sum_{p=1}^{r-1}\frac{1}{r-p}\frac{1}{z_{im}^{r-p}z_{nm}^p}\right.$$
$$\left.+\frac{\log(|z_{im}|^2/|z_{\rm in}|^2)}{z_{nm}^r}\right). \quad (C18)$$

---

[7]When computing this integral, one will yield anomalous expressions that spoil rotational invariance, more specifically expressions containing $(\bar{z}_{\rm in}-z_{mn})$.

Following the same procedures as above, we find the $s=1$, $t>1$ integrals to be

$$\frac{1}{\pi}\int d^2x\frac{\log(\mu^2|z-z_n|^2)}{(z-z_m)^{r+1}(\bar{z}-\bar{z}_i)^t}$$
$$=-\frac{1}{r(t-1)}\left(\frac{1}{z_{nm}^r\bar{z}_{ni}^{t-1}}-\frac{1}{\bar{z}_{ni}^{t-1}z_{im}^r}-\frac{1}{z_{nm}^r\bar{z}_{mi}^{t-1}}\right). \quad (C19)$$

For the integrals with $s>1$, $t\geq 1$, performing the reduction as in (C14) gives

$$\frac{1}{\pi}\int\frac{d^2x}{(z-z_m)^{r+1}(\bar{z}-\bar{z}_n)^{s-1}(\bar{z}-\bar{z}_i)^t}$$
$$=-\frac{1}{r}\binom{t+s-3}{s-2}\frac{1}{\bar{z}_{\rm in}^{s-1}\bar{z}_{ni}^{t-1}}\left(\frac{1}{z_{im}^r}-\frac{1}{z_{nm}^r}\right). \quad (C20)$$

We have thus completed the evaluation of all of the integrals in (C8).

[1] A. B. Zamolodchikov, Expectation value of composite field $T\bar{T}$ in two-dimensional quantum field theory, arXiv:hep-th/0401146.
[2] F. A. Smirnov and A. B. Zamolodchikov, On space of integrable quantum field theories, Nucl. Phys. **B915**, 363 (2017).
[3] A. Cavaglià, S. Negro, I. M. Szécsényi, and R. Tateo, $T\bar{T}$-deformed 2D quantum field theories, J. High Energy Phys. 10 (2016) 112.
[4] V. Rosenhaus and M. Smolkin, Integrability and renormalization under $T\bar{T}$, Phys. Rev. D **102**, 065009 (2020).
[5] B. Le Floch and M. Mezei, KdV charges in $T\bar{T}$-theories and new models with super-Hagedorn behavior, SciPost Phys. **7**, 043 (2019).
[6] G. Jorjadze and S. Theisen, Canonical maps and integrability in $T\bar{T}$ deformed 2d CFTs, arXiv:2001.03563.
[7] M. Guica, An integrable Lorentz-breaking deformation of two-dimensional CFTs, SciPost Phys. **5**, 048 (2018).
[8] N. Callebaut, J. Kruthoff, and H. Verlinde, $T\bar{T}$ deformed CFT as a non-critical string, J. High Energy Phys. 04 (2020) 084.
[9] A. J. Tolley, $T\bar{T}$ deformations, massive gravity and non-critical strings, J. High Energy Phys. 06 (2020) 050.
[10] J. Cardy, The $T\bar{T}$ deformation of quantum field theory as random geometry, J. High Energy Phys. 10 (2018) 186.
[11] S. Hirano and M. Shigemori, Random boundary geometry and gravity dual of $T\bar{T}$ deformation, J. High Energy Phys. 11 (2020) 108.
[12] S. Dubovsky, V. Gorbenko, and M. Mirbabayi, Asymptotic fragility, near AdS$_2$ holography and $T\bar{T}$, J. High Energy Phys. 09 (2017) 136.
[13] S. Dubovsky, V. Gorbenko, and G. Hernández-Chifflet, $T\bar{T}$ partition function from topological gravity, J. High Energy Phys. 09 (2018) 158.
[14] L. McGough, M. Mezei, and H. Verlinde, Moving the CFT into the bulk with $T\bar{T}$, J. High Energy Phys. 04 (2018) 010.
[15] P. Kraus, J. Liu, and D. Marolf, Cutoff AdS$_3$ versus the $T\bar{T}$ deformation, J. High Energy Phys. 07 (2018) 027.
[16] T. Hartman, J. Kruthoff, E. Shaghoulian, and A. Tajdini, Holography at finite cutoff with a $T^2$ deformation, J. High Energy Phys. 03 (2019) 004.
[17] P. Caputa, S. Datta, and V. Shyam, Sphere partition functions & cut-off AdS, J. High Energy Phys. 05 (2019) 112.
[18] W. Donnelly and V. Shyam, Entanglement entropy and $T\bar{T}$ deformation, Phys. Rev. Lett. **121**, 131602 (2018).
[19] B. Chen, L. Chen, and P. X. Hao, Entanglement entropy in $T\bar{T}$-deformed CFT, Phys. Rev. D **98**, 086025 (2018).
[20] Y. Li and Y. Zhou, Cutoff AdS$_3$ versus $T\bar{T}$ CFT$_2$ in the large central charge sector: Correlators of energy-momentum tensor, J. High Energy Phys. 12 (2020) 168.

[21] S. He, Y. Li, Y. Z. Li, and Y. Zhang, Holographic torus correlators of stress tensor in AdS$_3$/CFT$_2$, J. High Energy Phys. 06 (2023) 116.

[22] A. Bzowski and M. Guica, The holographic interpretation of $J\bar{T}$-deformed CFTs, J. High Energy Phys. 01 (2019) 198.

[23] M. Guica and R. Monten, $T\bar{T}$ and the mirage of a bulk cutoff, SciPost Phys. **10**, 024 (2021).

[24] S. Datta and Y. Jiang, $T\bar{T}$-deformed partition functions, J. High Energy Phys. 08 (2018) 106.

[25] O. Aharony, S. Datta, A. Giveon, Y. Jiang, and D. Kutasov, Modular invariance and uniqueness of $T\bar{T}$-deformed CFT, J. High Energy Phys. 01 (2019) 086.

[26] J. Cardy, $T\bar{T}$-deformed modular forms, Commun. Num. Theor. Phys. **16**, 435 (2022).

[27] M. Asrat, KdV charges and the generalized torus partition sum in $T\bar{T}$ deformation, Nucl. Phys. **B958**, 115119 (2020).

[28] S. He, Y. Sun, and Y. X. Zhang, $T\bar{T}$-flow effects on torus partition functions, J. High Energy Phys. 09 (2021) 061.

[29] S. Hirano, T. Nakajima, and M. Shigemori, $T\bar{T}$ Deformation of stress-tensor correlators from random geometry, J. High Energy Phys. 04 (2021) 270.

[30] Y. Li, Comments on large central charge $T\bar{T}$-deformed conformal field theory and cutoff AdS holography, arXiv:2012.14414.

[31] S. Ebert, E. Hijano, P. Kraus, R. Monten, and R. M. Myers, Field theory of interacting boundary gravitons, SciPost Phys. **13**, 038 (2022).

[32] S. He, J. R. Sun, and Y. Sun, The correlation function of (1,1) and (2,2) supersymmetric theories with $T\bar{T}$-deformation, J. High Energy Phys. 04 (2020) 100.

[33] S. He and H. Shu, Correlation functions, entanglement and chaos in the $T\bar{T}/J\bar{T}$-deformed CFTs, J. High Energy Phys. 02 (2020) 088.

[34] S. He, Note on higher-point correlators of the $T\bar{T}$ or $J\bar{T}$ deformed CFTs, Sci. China Phys. Mech. Astron. **64**, 291011 (2021).

[35] S. Ebert, H. Y. Sun, and Z. Sun, $T\bar{T}$ deformation in SCFTs and integrable supersymmetric theories, J. High Energy Phys. 09 (2021) 082.

[36] S. He and Y. Z. Li, Genus two correlation functions in CFTs with $T\bar{T}$ deformation, Sci. China Phys. Mech. Astron. **66**, 251011 (2023).

[37] S. He, J. Yang, Y. X. Zhang, and Z. X. Zhao, Pseudo entropy of primary operators in $T\bar{T}/J\bar{T}$-deformed CFTs, J. High Energy Phys. 09 (2023) 025.

[38] M. Guica and R. Monten, Infinite pseudo-conformal symmetries of classical $T\bar{T}$, $J\bar{T}$ and $JT_a$—deformed CFTs, SciPost Phys. **11**, 078 (2021).

[39] P. Kraus, R. Monten, and R. M. Myers, 3D gravity in a box, SciPost Phys. **11**, 070 (2021).

[40] M. He, S. He, and Y. h. Gao, Surface charges in Chern-Simons gravity with $T\bar{T}$ deformation, J. High Energy Phys. 03 (2022) 044.

[41] A. Dey and A. Fortinsky, Perturbative renormalization of the $T\bar{T}$-deformed free massive Dirac fermion, J. High Energy Phys. 12 (2021) 200.

[42] J. Cardy, $T\bar{T}$ deformation of cerrelation functions, J. High Energy Phys. 12 (2019) 160.

[43] O. Aharony and N. Barel, Correlation functions in $T\bar{T}$-deformed Conformal Field Theories, J. High Energy Phys. 08 (2023) 035.

[44] J. Liu, J. Haruna, and M. Yamada, Non-perturbative aspects of two-dimensional $T\bar{T}$-deformed scalar theory from functional renormalization group, Phys. Rev. D **109**, 065016 (2024).

[45] W. Cui, H. Shu, W. Song, and J. Wang, Correlation functions in the TsT/$T\bar{T}$ correspondence, J. High Energy Phys. 04 (2024) 017.

[46] S. B. Chakraborty, S. Georgescu, and M. Guica, States, symmetries and correlators of $T\bar{T}$-and $J\bar{T}$ symmetric orbifolds, SciPost Phys. **16**, 011 (2024).

[47] J. Kruthoff and O. Parrikar, On the flow of states under $T\bar{T}$, arXiv:2006.03054.

[48] G. Bonelli, N. Doroud, and M. Zhu, $T\bar{T}$-deformations in closed form, J. High Energy Phys. 06 (2018) 149.

[49] S. Hirano and M. Shigemori, Conformal field theory on $T\bar{T}$-deformed space and correlators from dynamical coordinate transformations, J. High Energy Phys. 07 (2024) 190.

[50] P. Kraus, R. Monten, and K. Roumpedakis, Refining the cutoff 3D gravity/$T\bar{T}$ correspondence, J. High Energy Phys. 10 (2022) 094.

[51] P. Di Francesco, P. Mathieu, and D. Senechal, *Conformal Field Theory* (Springer-Verlag, Berlin, 1997), ISBN 978-0-387-94785-3, 978-1-4612-7475-9.

[52] S. He, Y. Z. Li, and Y. Zhang, Holographic torus correlators in AdS$_3$ gravity coupled to scalar field, J. High Energy Phys. 05 (2024) 254.